Azide modification of single-walled carbon nanotubes for near-infrared defect photoluminescence by luminescent sp$^2$ defect formation

Keita Hayashi,[a] Yoshiaki Niidome,[a] Tamehito Shiga,[a] Boda Yu,[a] Yasuto Nakagawa,[a] Dawid Janas,[b] Tsuyohiko Fujigaya,[*a,c,d] and Tomohiro Shiraki[*a,c]

[a] *Department of Applied Chemistry, Kyushu University, 744 Motooka, Nishi-ku, Fukuoka, 819-0395, Japan, E-mail: fujigaya.tsuyohiko.948@m.kyushu-u.ac.jp; shiraki.tomohiro.992@m.kyushu-u.ac.jp.*
[b] *Department of Organic Chemistry, Bioorganic Chemistry and Biotechnology, Silesian University of Technology, B. Krzywoustego 4, 44-100, Gliwice, Poland*
[c] *International Institute for Carbon-Neutral Energy Research (WPI-I2CNER), Kyushu University, 744 Motooka, Nishi-ku, Fukuoka, 819-0395, Japan.*
[d] *Center for Molecular Systems (CMS), Kyushu University, 744 Motooka, Nishi-ku, Fukuoka, 819-0395, Japan.*

Azide functionalization produced luminescent sp$^2$-type defects on single-walled carbon nanotubes, by which defect photoluminescence appeared in near infrared regions (1116 nm). Changes in exciton properties were induced by localization effects at the defect sites, creating exciton-engineered nanomaterials based on the defect structure design.

Chemical functionalization of carbon nanotubes (CNTs) via covalent attachment of various molecules and polymers has been widely used to both solubilize CNTs in various solvents and to tailor their surfaces.[1] For single-walled CNTs (SWCNTs), a limited amount of chemical modification in terms of local chemical functionalization has been reported;[2-7] It has been used to dope local structural defects in the SWCNT sp$^2$ carbon networks via covalent bond formation between the nanotube surfaces and the modifier molecules. These locally functionalized SWCNTs (lf-SWCNTs) show a new defect near-infrared (NIR) photoluminescence ($E_{11}$* PL) from the defect-doped sites.[2-7] The $E_{11}$* PL is red-shifted and bright relative to the intrinsic PL ($E_{11}$ PL) of pristine SWCNTs, because the defect doping reduces the bandgaps and localize photogenerated excitons at the defect sites for the efficient PL generation. Sp$^3$-carbon defects have been generally produced using diazonium chemistry, reductive alkylation, and so on. These organic chemistry approaches have enabled the attachment of functional molecules on defect-doped sites for dynamic $E_{11}$* PL wavelength variations based on photochromic reactions[8] and molecular recognition[9, 10], as well as for broad $E_{11}$* PL wavelength modulation.[11-13] Properties of the localized excitons in lf-SWCNTs could be significantly changed depending on the structures of the defect-doped sites.[14] Moreover, new chemical functionalization techniques have contributed to the formation of defect-doped sites with various molecular structures and defect arrangements for lf-SWCNT.[15-17] Thus, further development of local chemical functionalization methods could enable broader chemical structures at lf-SWCNTs defect-doped sites for the creation of novel defect PL properties including the enhancement of $E_{11}$* PL wavelengths and enhanced PL quantum yields.

Azide compounds have been used for surface functionalization of SWCNTs (Fig. S1). The azide group is activated to produce a reactive nitrene and its [2+1] cycloaddition occurs on the SWCNT surface. Subsequent C-C bond cleavage results in the formation of two $sp^2$ carbon atoms on the SWCNT, which bind with the nitrogen moiety of the functionalized molecule.[18] This $sp^2$ carbon linkage could partially maintain the π-conjugation of the SWCNTs, but was not examined as a luminescent defect in lf-SWCNT synthesis. However, the non-planar strained $sp^2$ bonding possibly offers $sp^2$-type defects that have been observed only in oxygen-doped lf-SWCNTs with ether-based structural defects.[19-21] Here, the $sp^2$-type defect formed by the reaction of SWCNTs and 4-azidobenzoic acid (BA-N$_3$) (see Fig. 1) works as a luminescent defect to emit $E_{11}$* PL from the product lf-SWCNTs-N. The exciton localized at the $sp^2$-type defect-doped sites in the lf-SWCNTs-N exhibits a PL solvatochromism different from that of typical lf-SWCNTs with $sp^3$-type defects. Moreover, further characterization of lf-SWCNTs-N revealed an exciton detrapping energy comparable to those of previous lf-SWCNTs and a stable PL wavelength with respect to pH changes.

lf-SWCNTs-N were synthesized by reacting SWCNTs [CoMoCAT, (6,5) chirality rich] solubilized in an aqueous sodium dodecyl sulfate (SDS) solution with BA-N$_3$ under ultraviolet (UV) light. Fig. 2 shows PL spectra of pristine SWCNTs, lf-SWCNTs-N, and lf-SWCNTs synthesized via diazonium chemistry using p-carboxybenzene tetrafluoroborate (BA-N$_2$BF$_4$) (lf-SWCNTs-C), by which the lf-SWCNT defect PL from $sp^2$- and $sp^3$-type defects were compared. The excitation wavelength was 570 nm, which corresponded to the second-lowest transition ($E_{22}$) in (6,5) SWCNTs. For pristine SWCNTs, a peak appeared at 980 nm, which was assigned to the PL signal based on the lowest energy transition ($E_{11}$) of (6,5) SWCNTs. A small peak at 1115 nm was assigned to the $E_{11}$ PL of (8,4) SWCNTs that slightly contaminated the used (6,5)-rich SWCNT sample. For lf-SWCNTs-N, additional PL was observed at 1116 nm. A small PL peak was also detected at 1233 nm, which could be attributed to the formation of a minor defect binding configuration caused by a different binding position of the functionalized molecule on the SWCNT surface.[5] In the PL spectra of lf-SWCNTs-N synthesized using different concentrations of BA-N$_3$ (Fig. S2), the intensities increased with BA-N$_3$ concentrations up to 0.8 mM. In contrast, 1.6-mM BA-N$_3$ reduced the PL intensity at 1116 nm, which may be attributed to a decreased reaction efficiency by side reactions, such as dimerization of nitrene intermediates at high concentration.[22] When the excitation wavelength was changed stepwise from 560 nm to 580 nm, no peak shifts were observed, indicating that the PL was emitted from lf-SWCNTs-N with (6,5) chirality (Fig. S3).[23] For the lf-SWCNTs-C,[24] $E_{11}$* PL was observed at 1128 nm, which was a longer wavelength than that for lf-SWCNTs-N (1116 nm). In the visible (vis)/NIR absorption spectra of lf-SWCNTs-N (Fig. S4), the $E_{11}$ absorption band of the (6,5)-chirality SWCNTs was observed at 980 nm, and its absorbance was mostly unchanged after reacting with BA-N$_3$. Hence, the π-conjugation of the sp$^2$ SWCNT network was mostly preserved for the sp$^2$-type defect structures, as reported previously.[18] In the Raman spectrum of lf-SWCNTs-N, a peak was observed at 1289 cm$^{-1}$, which was assigned to the D band relating to sp$^3$ defect structures.[25] Its intensity was unchanged relative to that of pristine SWCNTs (Fig. S5). This was also consistent with the reported observation of sp$^2$-type defect formation in azide functionalization of SWCNTs.[26] In contrast, lf-SWCNTs-C exhibited an increase in the D-band intensity because of the introduction of sp$^3$-type defects by the diazonium functionalization.[13] In the spectrum of X-ray photoelectron spectroscopy (XPS)of lf-SWCNTs-N (Fig. S6), a clear peak was detected at 400.1 eV for the N 1s orbital, indicating a sp$^2$-type defect structure formation by the covalent modification using BA-N$_3$.[26] Here, a UV lamp excited a broad wavelength range over 300–400 nm, and its overlap with the absorption band edge of BA-N$_3$ induced the photoactivated functionalization (Fig.S7). Removal of dissolved oxygen in the reaction solutions was necessary to proceed with this functionalization because a lack of N$_2$ gas bubbling treatment prevented 1116-nm PL generation by quenching the photogenerated nitrene species.[27] Therefore, the chemical functionalization of SWCNTs using BA-N$_3$ created sp$^2$-type defect structures in lf-SWCNTs-N, and $E_{11}$* PL defect emission was observed at 1116 nm.

Fig. 3 shows PL spectra of lf-SWCNTs synthesized using azide compounds 2 and 3, respectively (lf-SWCNTs-N2 and lf-SWCNTs-N3). The lf-SWCNTs-N2 and lf-SWCNTs-N3 exhibited $E_{11}$* PL at 1113 nm and 1098 nm, respectively. Thus, the chemical structural differences of the aryl moieties in the azide compounds changed the $E_{11}$* PL wavelength of lf-SWCNTs with sp$^2$-type defects, as similarly observed for lf-SWCNTs synthesized via diazonium chemistry.[2, 3, 7] Therefore, molecular structures of azide reactants could modulate $E_{11}$* PL wavelengths and bind various functional molecules at sp$^2$-type defects on lf-SWCNTs, offering another series of molecularly functionalized lf-SWCNTs.

$E_{11}$* PL is emitted from excitons localized at the defect dopes sites of lf-SWCNTs. Accordingly, the exciton localization states are important factors in determining the localized exciton properties and the $E_{11}$* PL.[14, 28] Here, we investigated the effects of the sp$^2$ and sp$^3$ defect structure differences in lf-SWCNTs on localized exciton properties by using a PL solvatochromism.[14, 28] Specifically, we injected water-immiscible organic solvents at the interface between the SWCNTs and surfactant micelles to create various dielectric environments by mixing an aqueous solution of sodium dodecylbenzene sulfonate (SDBS)-coated lf-SWCNTs and the organic solvents (toluene, o-xylene, 2,6-dichlorotoluene, 3,4-dichlorotoluene, or o-dichlorobenzene). Prior to the experiments, the coating surfactants were changed from SDS to SDBS by adding a 1.0-wt% SDBS micellar solution to the lf-SWCNTs-N with a SDS coating; this was needed for stable SDBS micelle coating to maintain the solubilized SWCNT states after organic solvent injection.[29] After injecting the organic solvents to lf-SWCNTs-N and lf-

SWCNTs-C, $E_{11}$* PL and $E_{11}$ PL were both red-shifted (Table S1) because of the relatively higher-polarity environment formation surrounding nanotubes. The observed PL red-shift phenomena were consistent with those reported for lf-SWCNTs synthesized via diazonium chemistry.[14, 28] Thus, as shown in Figs. 4 and S8, the observed PL energy shifts $\Delta E_{11}$* and $\Delta E_{11}$ were plotted as a function of the orientation polarity parameter $f(\varepsilon) - f(\eta^2) = 2(\varepsilon - 1)/(2\varepsilon + 1) - 2(\eta^2 - 1)/(2\eta^2 + 1)$,[30] where $\varepsilon$ and $\eta$ are dielectric constant and the refractive index of the solvent, respectively. For the solvatochromic PL energy shifts, dielectric exciton-solvent interactions resulting from the dipole moments and polarizabilities were considered.[30-32] Therefore, a linear combination of an exciton dipole-solvent polarizability term, an exciton polarizability-solvent dipole term, and a polarizability-polarizability term were included in Eq. (1).[14] A dipole–dipole interaction term was excluded because it was expected to have a negligible effect because of the large time-scale difference between radiative relaxation of the excitonic PL(~100 ps)[33, 34] and the reorientation relaxation of the tightly bound solvent molecules on the SWCNT surface.[35, 36]

$$\Delta E = -L_1 \frac{\Delta \alpha}{R^3} \Delta[f(\varepsilon) - f(\eta^2)] - \frac{1}{R^3}(L_2 \Delta \alpha + \frac{\Delta(\mu^2)}{2}) \Delta f(\eta^2). \qquad (1)$$

In Eq. (1), $f(\eta^2) = 2(\eta^2 - 1)/(2\eta^2 + 1)$ is an induction polarity parameter, $L_1$ and $L_2$ are fluctuation factors, $\Delta\alpha$ is a polarizability change, $\Delta(\mu^2)$ is a change in the square value of the dipole moment, and $R$ is an effective radius. Given that the fluctuation factors and $R$ will be the same, the slopes in Fig. 4 correspond to the first term in Eq. (1), and the observed similar slopes for $\Delta E_{11}$* in lf-SWCNTs-N and lf-SWCNTs-C result in mostly same $\Delta\alpha$ at their defect-doped sites. The intercepts in Fig. 4 correspond to the second term in Eq. (1), and only $\Delta(\mu^2)$ changes the intercept values. Thus, the smaller intercept value for lf-SWCNTs-N was attributed to the decreased dipole moment of the localized exciton at the $sp^2$-type defect relative to that at $sp^3$ type defect. This would be because of less exciton localization in the preservation of the π-conjugated carbon network for the $sp^2$-type defect structure. In contrast, as shown in Fig. S8, similar slopes and intercept values were observed for $E_{11}$ PL from lf-SWCNTs-N and lf-SWCNTs-C because it originated from mobile excitons on the pristine sites of the tubes. Consequently, the defect–structure differences based on modifier molecules could modulate localized exciton properties and $E_{11}$* PL solvatochromism behaviors of lf-SWCNTs.

For the $sp^2$-type defects in lf-SWCNTs-N, the energy for exciton detrapping, which is a phonon-assisting delocalization process of the localized excitons, was evaluated based on Van't Hoff plot analysis (Fig. S9).[37] The exciton detrapping energy for lf-SWCNTs-N was estimated to be 100 meV, which was similar to that for lf-SWCNTs-C (97.1 meV).

Because BA-N$_3$ has a carboxylic group, pH effects on the $E_{11}$* PL wavelength of lf-SWCNTs-N were investigated. Lf-SWCNTs with $sp^3$-type defects synthesized via diazonium chemistry exhibited $E_{11}$* PL shifts based on protonation and deprotonation of amine substituents of the functionalized aryl groups.[10, 38] Here, considering a p$K$a value (~4.2) for benzoic acid,[39] the pH was adjusted to 2.6 and

9.1. For the lf-SWCNTs-N at different pH conditions (Fig. 5), no $E_{11}$* PL spectral shifts were observed. In contrast, lf-SWCNTs-C exhibited 7.5-nm shifts from the pH difference. The weaker responsiveness of the lf-SWCNTs-N may be attributed to structural differences of the $sp^2$- and $sp^3$-type defects. That is, lf-SWCNTs-N have benzoic acid groups with nitrogen linkages on $sp^2$ carbons, which was different from the direct aryl-group attachment for the $sp^3$ defect sites. Lf-SWCNTs-N2 and -N3 also exhibited no $E_{11}$* PL spectral changes with pH variation (Fig. S10). Therefore, lf-SWCNTs-N had stable $E_{11}$* PL wavelengths, regardless of the surrounding environment differences such as pH.

In summary, we synthesized lf-SWCNTs-N with $sp^2$ defect-doped sites by the reaction with photo-activated BA-$N_3$. The $sp^2$ defect structure formation was confirmed with XPS and Raman spectroscopy. The lf-SWCNTs-N emitted $E_{11}$* PL at 1116 nm, which was a shorter wavelength than that of lf-SWCNTs-C that tethered the same benzoic acid groups on $sp^3$ defect-doped sites. From the PL solvatochromism analysis, it is found that exciton property changes occurred depending on the defect structural differences of the lf-SWCNTs. Namely, the dipole moment of the localized exciton at $sp^2$ defect sites in lf-SWCNTs-N decreased because of less exciton localization in the preserved π-conjugation of the $sp^2$ defect structure. In addition, the lf-SWCNTs-N had a 100-meV exciton detrapping energy and no pH-induced spectral shifts. Azide compounds have been widely used for chemical functionalization of carbon nanotubes, but this is the first observation of $E_{11}$* PL from the functionalized SWCNTs by this method. Exciton property variations based on $sp^2$-type defect formation enable engineered localized excitons in lf-SWCNTs for exciton-based quantum technologies. Therefore, the myriad of azide compounds could enable syntheses of new lf-SWCNTs with $sp^2$ defect-doped sites for advanced NIR applications of PL wavelengths longer than 1000 nm, such as quantum telecom technologies and deep-tissue bioimaging.

We thank for financial supports of Grant-in-Aid for JSPS KAKENHI Grant Number JP19H02557 and JP22H01910, UMO-2020/39/D/ST5/00285 from National Science Centre, Poland, and the Nanotechnology Platform Project from MEXT, Japan.

Conflicts of interest:   There are no conflicts to declare.

Notes and references

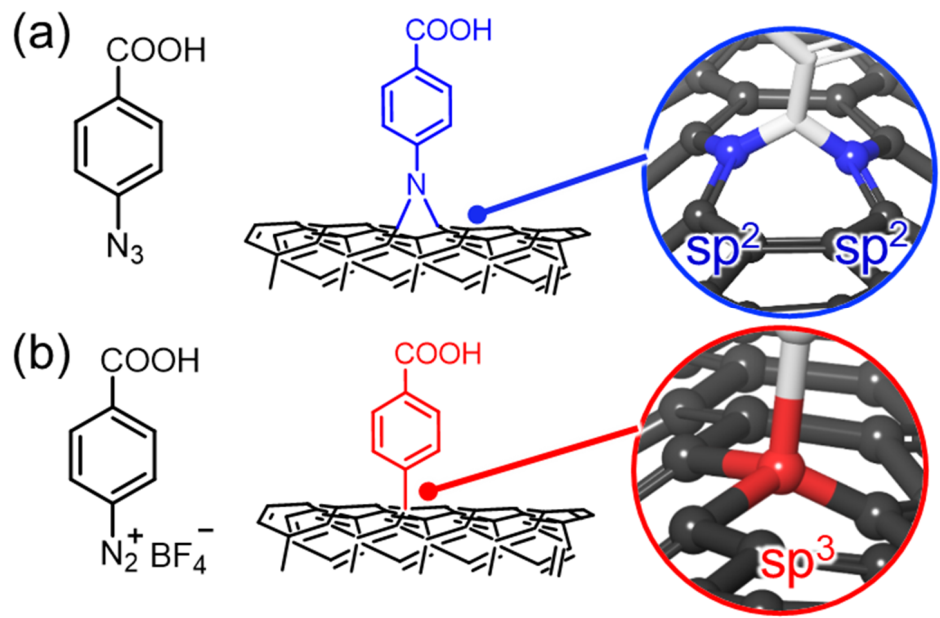

**Fig. 1** (a) Chemical structure of BA-N$_3$ and its functionalization to form sp$^2$-type defect structure in lf-SWCNTs-N and (b) chemical structure of BA-N$_2$BF$_4$ and its functionalization to form sp$^3$-type defect structure in lf-SWCNTs-C.

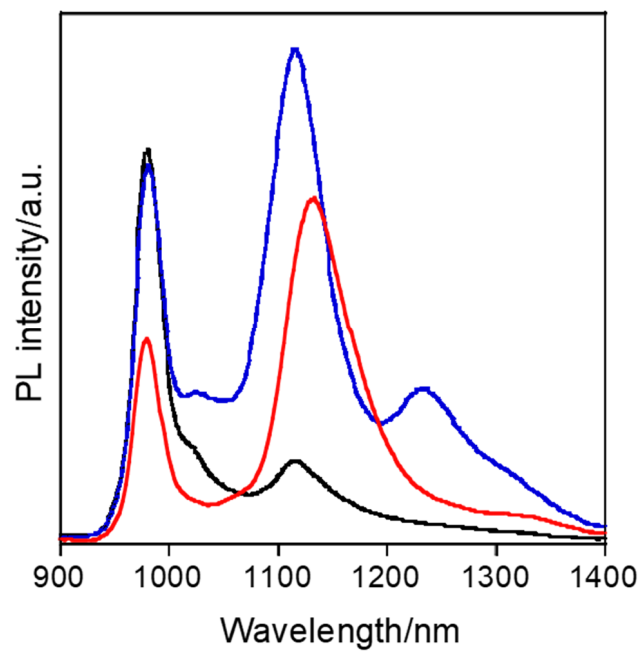

**Fig. 2** PL spectra of pristine SWCNTs (black), lf-SWCNTs-N (blue), and lf-SWCNTs-C (red). (570-nm excitation).

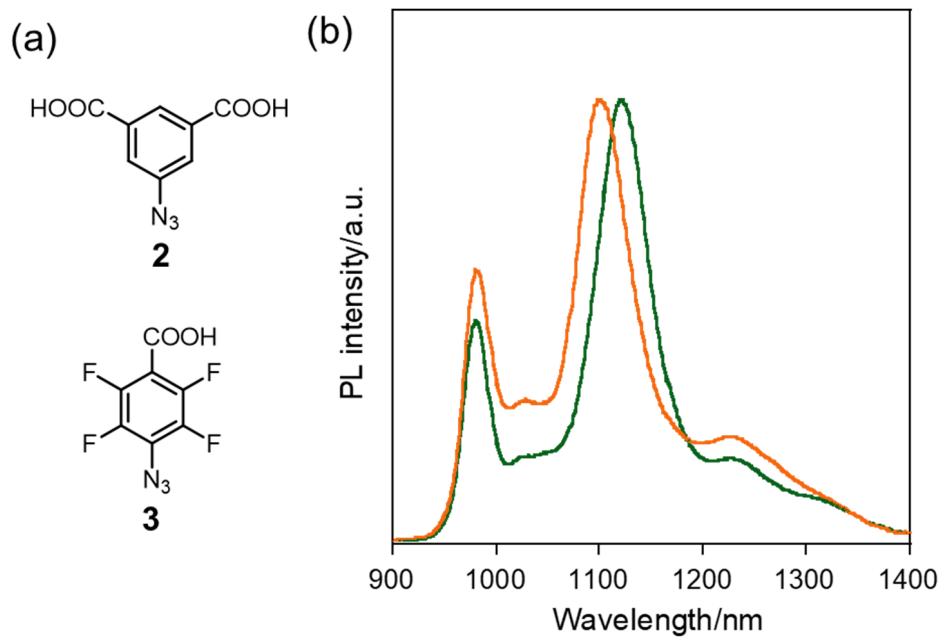

**Fig. 3** (a) Chemical structures of azide compounds **2** and **3**. (b) PL spectra of lf-SWCNTs-N2 (green) and lf-SWCNTs-N3 (orange) synthesized using 0.8-mM solutions of **2** and **3**, respectively. (570-nm excitation).

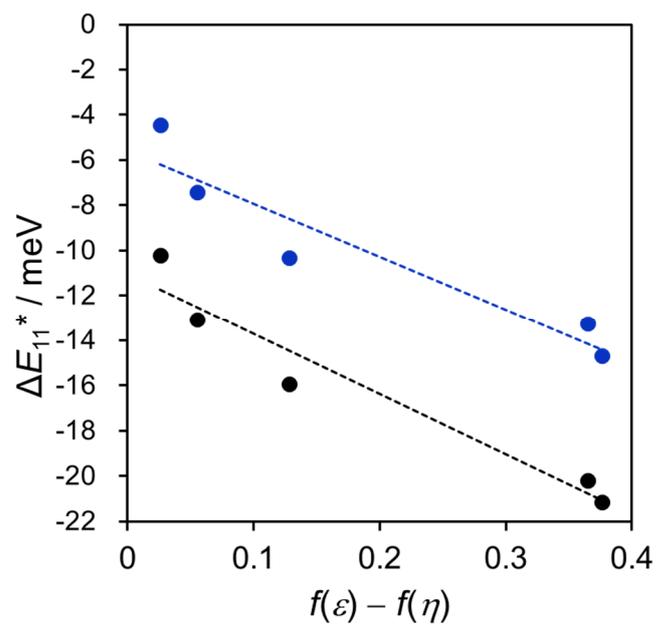

**Fig. 4** Plot of $\Delta E_{11}^*$ for lf-SWCNTs-N (blue) and lf-SWCNTs-C (black) as a function of $f(\varepsilon) - f(\eta^2)$ of the injected solvents. The dotted lines were obtained by a linear approximation.

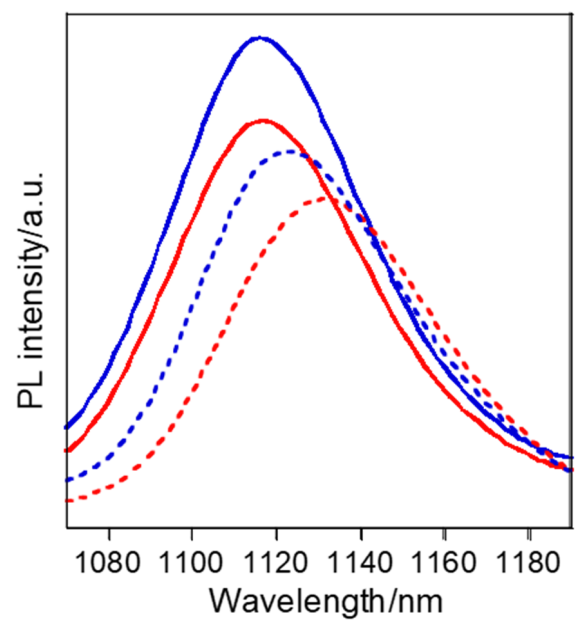

**Fig. 5** PL spectra of lf-SWCNTs-N (solid line) and lf-SWCNTs-C (dashed line) at pH 9.1 (blue) and 2.6 (red). (570-nm excitation).

# Azide modification of single-walled carbon nanotubes for near-infrared defect photoluminescence by luminescent sp² defect formation

Keita Hayashi,[a] Yoshiaki Niidome,[a] Tamehito Shiga,[a] Boda Yu,[a] Yasuto Nakagawa,[a] Dawid Janas,[b] Tsuyohiko Fujigaya,[*a,c,d] and Tomohiro Shiraki[*a,c]

[a] *Department of Applied Chemistry, Kyushu University, 744 Motooka, Nishi-ku, Fukuoka, 819-0395, Japan, E-mail: fujigaya.tsuyohiko.948@m.kyushu-u.ac.jp; shiraki.tomohiro.992@m.kyushu-u.ac.jp.*
[b] *Department of Organic Chemistry, Bioorganic Chemistry and Biotechnology, Silesian University of Technology, B. Krzywoustego 4, 44-100, Gliwice, Poland*
[c] *International Institute for Carbon-Neutral Energy Research (WPI-I2CNER) Kyushu University,744 Motooka, Nishi-ku, Fukuoka, 819-0395, Japan.*
[d] *Center for Molecular Systems (CMS), Kyushu University, 744 Motooka, Nishi-ku, Fukuoka, 819-0395, Japan.*

E-mail: fujigaya.tsuyohiko.948@m.kyushu-u.ac.jp; shiraki.tomohiro.992@m.kyushu-u.ac.jp

**Materials**

Sodium dodecyl benzene sulfonate (SDBS), 4-azidobenzoic acid (BA-N$_3$), 4-aminobenzoic acid, *o*-xylene, 2,6-dichlorotoluene, 3,4-dichlorotoluene, *o*-dichlorobenzene, 5-aminoisophthalic acid, and 4-azido-2,3,5,6-tetrafluorobenzoic acid (azide compound **3**) were purchased from the Tokyo Chemical Industry Co. Single-walled carbon nanotubes (SWCNTs) (CoMoCAT (6, 5) rich), deuterium oxide (D$_2$O), sodium dodecyl sulfate (SDS), toluene, and urea were purchased from the Sigma-Aldrich Co. Ethanol, tetrafluoroboric acid (48 wt.% aqueous solution), sodium nitrite, *aq*. HCl (36%), and NaN$_3$ were obtained from Wako Pure Chemical Industries. DCl and NaOD solutions (Sigma Aldrich) were used to adjust pH of the lf-SWCNT solutions. All chemicals were used as received.

**Instruments**

The $^1$H, $^{13}$C, $^{19}$F NMR spectra were recorded using JNM-ECZ400 from JEOL Ltd. The UV/vis/NIR and PL spectra were measured using a V-670 (JASCO) and a HORIBA JOBIN YVON spectrofluorometer (FluorologR-3 with FluorEssence), respectively. Fourier-transform infrared (FTIR) spectra were recorded using a Spectrum 65 (Perkin Elmer). The Raman spectra at an excitation of λ = 532 nm were recorded by a RAMAN touch spectrometer (Nanophoton Corporation), in which functionalized SWCNTs were collected by filtration and washed with water and ethanol , and then dried in vacuo for measurement samples. X-ray photoelectron spectroscopy spectra were measured by Kratos AXIS-ULTRA DLD (Shimadzu Corporation). For preparation of the SWCNT solutions, a bath-type sonicator (BRANSON, CPX5800H-J), a tip-type sonicator (Tomy Seiko, UD-211) and an ultracentrifuge (Hitachi, himac CS GXL) were used. A UV lamp (AS ONE, SLUV-4) was used for chemical functionalization of SWCNT using azide compounds. For pH measurements, a pH meter (B-712, Horiba) was used.

**Synthesis**

<u>*p*-Carboxybenzenediazonium tetrafluoroborate (BA-N$_2$BF$_4$)[S1]</u>

In a 10 mL two-necked flask, 273.4 mg (2.00 mmol) of 4-aminobenzoic acid and 1000 μL of water were added. Under a nitrogen atmosphere, the solution was cooled in an ice bath and 800 μL of *aq*. 42% tetrafluoroboric acid was added, and then stirred for 10 min. After adding 800 μL of an aqueous solution of NaNO$_2$ (154 mg, 2.23 mmol) dropwise, the reaction mixture was stirred for 20 min. The generated solid was collected by filtration and washed with diethyl ether. The product was dried under vacuum to yield a colorless powder (82.8 mg).

Yield 17.5%; $^1$H NMR (400 MHz, DMSO-$d_6$), δ/ppm = 8.74 (d, J = 8.7 Hz, 2H), 8.38 (d, J = 8.7 Hz, 2H); $^{19}$F NMR (400 MHz, , DMSO-$d_6$), δ/ppm = –148; FT-IR, ν/cm$^{-1}$ = 2309 (N≡N).

5-Azidoisophthalic acid (azide compound **2**) [S2]

In a 25 mL flask, 90.57 mg (0.500 mmol) of aminoisophthalic acid was dissolved in 5 mL of 18.5% HCl aqueous solution. After cooling the solution in an ice bath, 2 mL of an aqueous solution of $NaNO_2$ (36.23 mg, 0.525 mmol) was added dropwise and stirred for 1 h. To the solution, urea (3.30 mg, 0.0549 mmol) was added, and 3 mL of aqueous solution of $NaN_3$ (34,13 mg, 0.525 mmol) was then added dropwise under vigorous stirring. The reaction mixture was stirred for 30 min in an ice bath, followed by stirring for 2 h at room temperature. Extraction of the product was conducted after adding diethyl ether. The obtained organic layer was washed with NaCl saturated water and dried over $Na_2SO_4$. Evaporation of the solvent provided a colorless solid (95.8 mg).

Yield 92.5%; $^1$H NMR (400 MHz, DMSO-$d_6$), δ/ppm = 8.24 (t, J = 1.4 Hz, 1H), 7.76 (d, J = 1.4 Hz, 2H); FT-IR, ν/cm$^{-1}$ = 2125 (-$N_3$).

lf-SWCNTs-N, -N2, and -N3

In a 50 mL glass bottle, 5 mg of the SWCNTs was dispersed in a $D_2O$ solution of SDS (2.0wt.%) and sonicated using a bath-type sonicator for 1 h and a tip-type sonicator for 30 min. The resulting dispersion was ultracentrifuged at 147,000 g for 4 h and the supernatant (top ~80%) was collected as a SWCNT solution. The solution was diluted with $D_2O$ to prepare the appropriate SWCNT concentrations for absorption and PL measurements. The final samples for measurements were SWCNT solution with 0.20wt.% SDS.

In a 6 mL glass bottle, BA-$N_3$ was dissolved in 3.0 mL of the SWCNT solution whose pH was adjusted to be ~9 by adding a NaOD solution. After an $N_2$ gas bubbling treatment, the solution was irradiated by UV light (4 mW/cm$^2$) for 90 min for the chemical functionalization.

lf-SWCNTs-C

In a 6 mL glass bottle, *p*-carboxybenzenediazonium tetrafluoroborate was dissolved in a 0.20wt% SDS aqueous solution and mixed with the SWCNT solution. Therein, pH of the solution was adjusted to be ~9 by adding a NaOD solution. The mixed solution was left in the dark for 6 days for the chemical functionalization.

**Exciton detrapping energy evaluation experiments**[S3]

To produce Van't Hoff plots, PL measurements of the lf-SWCNTs-N and -C solutions were conducted at different temperature from 15 to 85 °C by using a temperature control unit. To maintain the solubilized SWCNT states in the applied temperatures, the coating surfactants were changed from SDS to SDBS by adding a 1.0 wt% SDBS micellar solution to the as-synthesized lf-SWCNTs solutions with SDS. Exciton detrapping energy ($\Delta E_T$) was estimated from the slopes of the plots using the following equation (S1).

$$\ln\left(\frac{I_{E_{11}}}{I_{E_{11}*}}\right) = \frac{\Delta E_\mathrm{T}}{kT} + A' \qquad (S1)$$

where $I_{E_{11}}$ and $I_{E_{11}*}$ are the integrated PL intensities of $E_{11}$ PL and $E_{11}*$ PL, $k$ is the Boltzmann constant, $T$ is temperature, and $A'$ is a correction factor. For this analysis, peak deconvolutions were conducted based on Lorenz functions.

**PL solvatochromism experiments**

To evaluate PL solvatochromism of the lf-SWCNTs-N and -C, the reported organic solvent injection method was applied for the creation of various dielectric environments.[S4-6] The used water-immiscible organic solvents are toluene, *o*-xylene, 2,6-dichlorotoluene, 3,4-dichlorotoluene, and *o*-dichlorobenzene. An organic solvent was added to the lf-SWCNTs solution (50/50 vol./vol.), and vigorously shaken using a vortex mixer (2000 rpm, 1.0 min), and then left until two phase separation of the organic and the aqueous layers occurred. The resultant aqueous layer was collected for measurements. By this treatment, the organic solvent was incorporated in the hydrophobic domain at the interface between the lf-SWCNTs and the SDBS-micelle coating.[S4, 7]

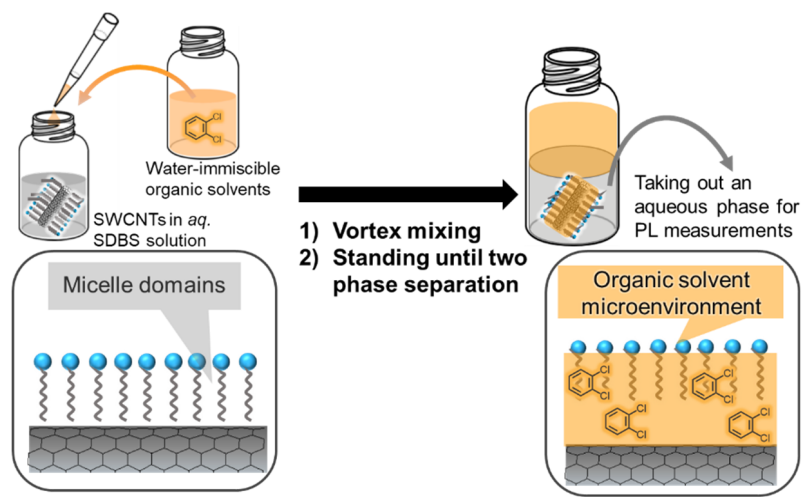

**Scheme. S1** Schematic of the experimental procedure of the organic solvent injection method for PL solvatochromism evaluations of the solubilized nanotubes.

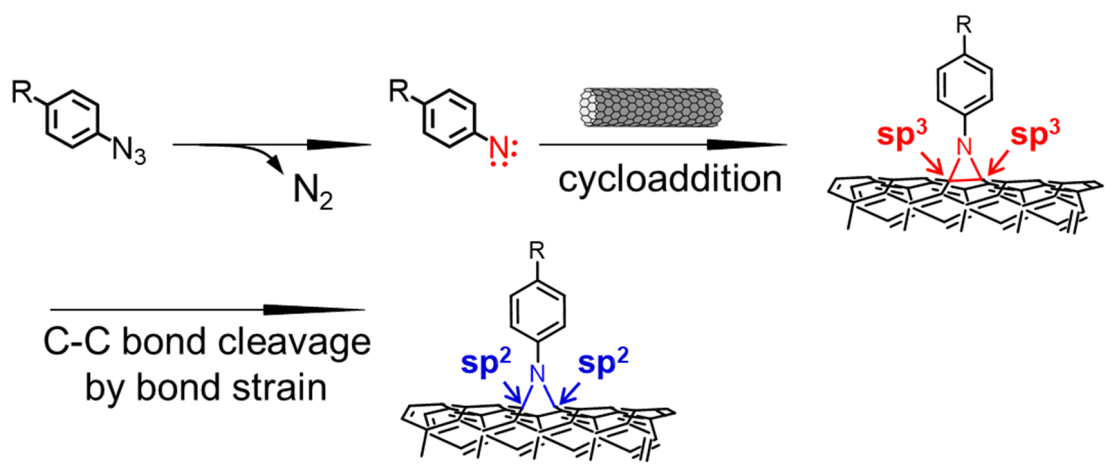

**Fig. S1** Proposed reaction scheme for SWCNT chemical functionalization using azide compouds.[S8]

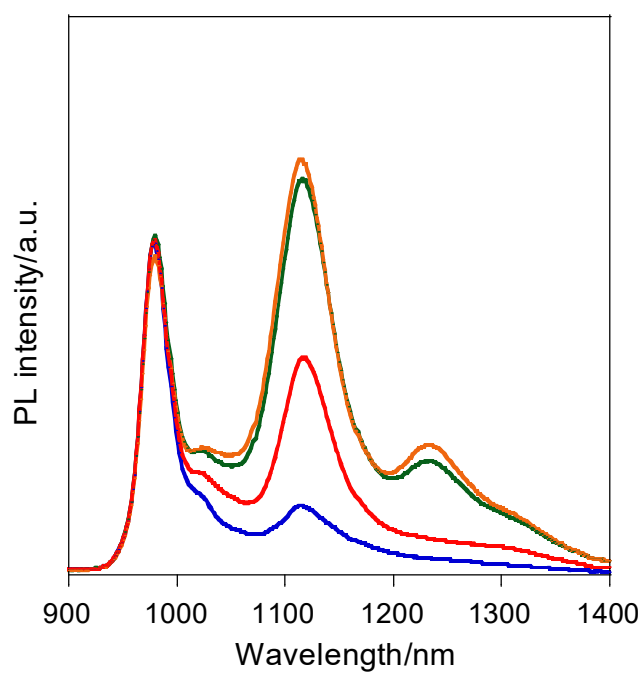

**Fig. S2** PL spectra of lf-SWCNTs-N synthesized using different concentrations of BA-N$_3$. [BA-N$_3$] = 0 (blue), 0.4 (green), 0.8 (orange), and 1.6 mM (red). $\lambda_{ex}$ = 570 nm.

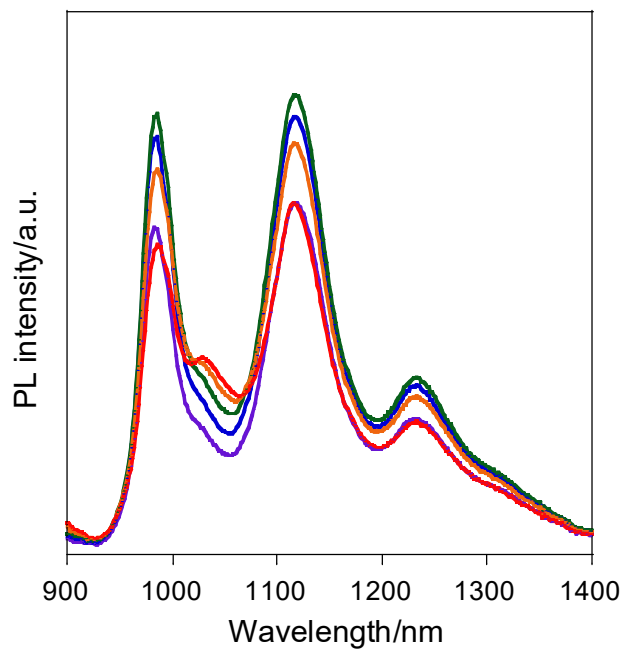

**Fig. S3** PL spectra of lf-SWCNTs-N collected by changing excitation wavelengths from 560 nm(purple) to 565 (blue), 570 (green), 575 (orange), and 580 nm (red).

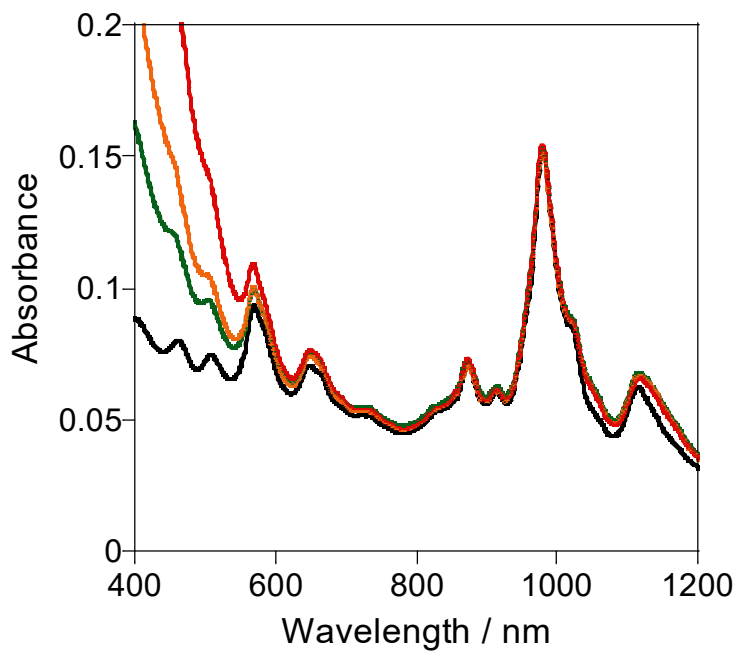

**Fig. S4** Vis/NIR absorption spectra of lf-SWCNTs-N synthesized using different concentrations of BA-N$_3$. [BA-N$_3$] = 0 (blue), 0.4 (green), 0.8 (orange), and 1.6 mM (red).

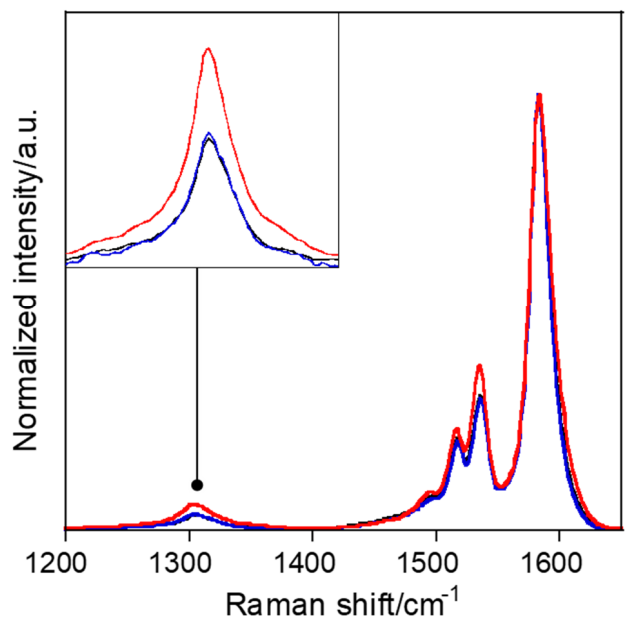

**Fig. S5** Raman spectra of pristine SWCNTs (black), lf-SWCNTs-N (blue), and lf-SWCNTs-C (red).

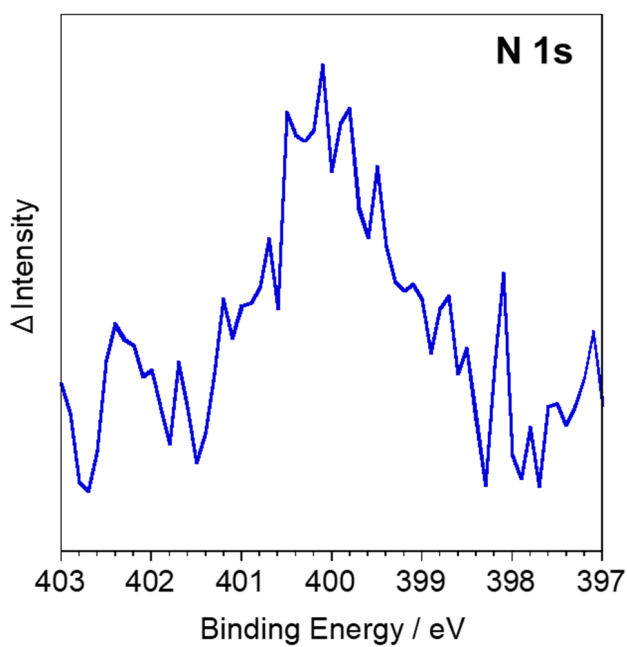

**Fig. S6** Differential XPS spectrum for N1s of lf-SWCNTs-N, which was obtained by subtracting the spectrum of pristine SWCNTs from that of lf-SWCNTs-N because of removal of a nitrogen signal from an impurity existing in the used SWCNT sample.

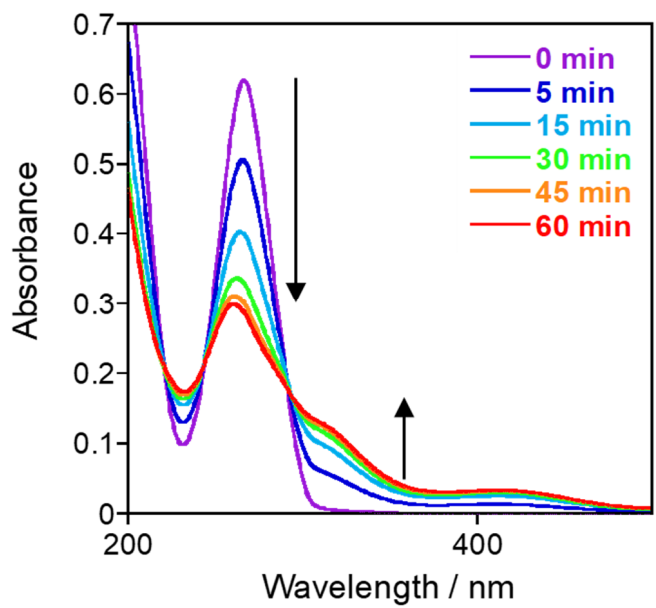

**Fig. S7** Time-course UV/vis absorption spectral changes of BA-N$_3$ in water by irradiating the UV light. The irradiation time was 0 (black), 5 (blue), 15 (sky blue), 30 (green), 45 (orange), 60 min (red).

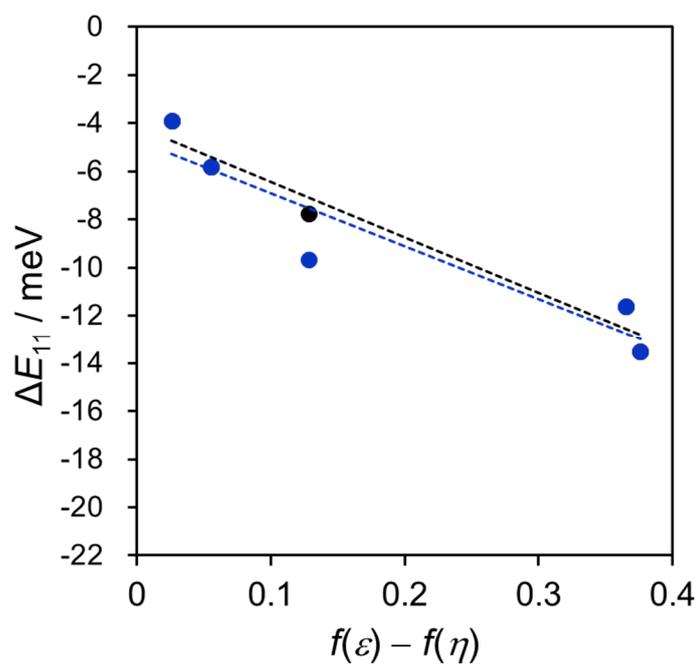

**Fig. S8** Plot of $\Delta E_{11}$ for lf-SWNTs-N (blue) and –C (black) as a function of $f(\varepsilon) - f(\eta^2)$ of the injected solvents. The dotted lines were obtained by a linear approximation method.

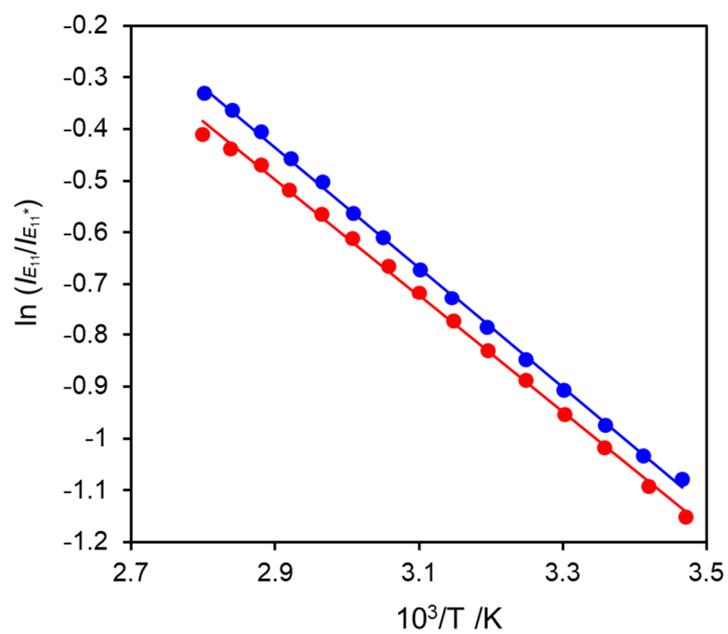

**Fig. S9** Van't Hoff plots for lf-SWCNTs-N (blue) and lf-SWCNTs-C (red).

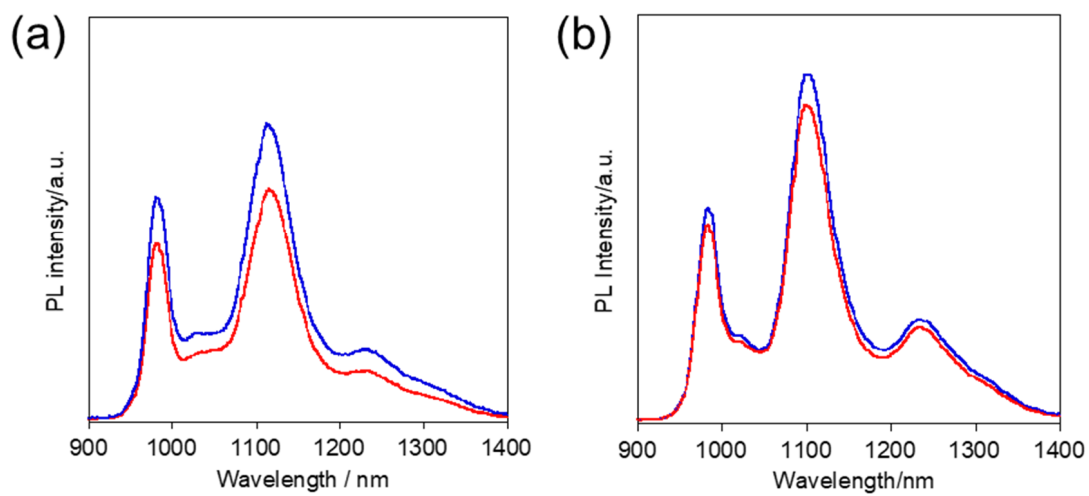

**Fig. S10** PL spectra of (a) lf-SWCNTs-N2 and (b) -N3 at pH 10 (blue) and 2.5 (red). $\lambda_{ex}$ = 570 nm.

**Table S1** Observed PL energy shifts of lf-SWCNTs-N and -C in organic solvent environments and their $f(\varepsilon) - f(\eta^2)$ values.

| environments | $f(\varepsilon) - f(\eta^2)$ | lf-SWCNTs-N | | | | lf-SWCNT-C | | | |
|---|---|---|---|---|---|---|---|---|---|
| | | $E_{11}$/nm | $\Delta E_{11}$/meV | $E_{11}^*$/nm | $\Delta E_{11}^*$/meV | $E_{11}$/nm | $\Delta E_{11}$/meV | $E_{11}^*$/nm | $\Delta E_{11}^*$/meV |
| SDBS micelle | - | 980.3 | 0.0 | 1116.0 | 0.0 | 980.3 | 0.0 | 1122.0 | 0.0 |
| toluene | 0.026 | 983.4 | -3.9 | 1123.5 | -4.5 | 983.4 | -3.9 | 1134.0 | -10.2 |
| o-xylene | 0.055 | 984.9 | -5.8 | 1126.5 | -7.4 | 984.9 | -5.8 | 1138.5 | -13.1 |
| 2,6-DCT | 0.128 | 987.9 | -9.7 | 1129.5 | -10.3 | 986.4 | -7.7 | 1143.0 | -15.9 |
| 3,4-DCT | 0.365 | 989.4 | -11.6 | 1131.0 | -13.2 | 989.4 | -11.6 | 1149.0 | -20.2 |
| oDCB | 0.376 | 990.9 | -13.5 | 1132.5 | -14.7 | 990.9 | -13.5 | 1150.5 | -21.1 |